\documentclass[aps,prl,twocolumn,groupedaddress,showpacs,preprintnumbers]{revtex4}
\usepackage{graphicx}
\usepackage{epstopdf}

\newcommand{\delete}[1]{}

\newcommand{\be}{\begin{equation}}
\newcommand{\ee}{\end{equation}}

\def\beq{\begin{equation}}
\def\eeq{\end{equation}}
\def\bea{\begin{eqnarray}}
\def\eea{\end{eqnarray}}
\def\ba{\begin{array}}
\def\ea{\end{array}}

\begin{document}

\preprint{CERN-PH-TH/2011-024}

\title{Phenomenology of TeV Little String Theory from Holography}


\author{Ignatios Antoniadis$^1$
}
\email[]{ignatios.antoniadis@cern.ch}
\thanks{On leave from CPHT (UMR CNRS 7644), Ecole Polytechnique, 91128 Palaiseau, France}
\author{Asimina Arvanitaki$^{2,3}$}
\email[]{aarvanitaki@lbl.gov}
\author{Savas Dimopoulos$^4$}
\email[]{savas@stanford.edu}
\author{Amit Giveon$^{5}$}
\email[]{giveon@phys.huji.ac.il}

\affiliation{$^1$Department of Physics, CERN - Theory Division, 1211 Geneva 23, Switzerland}
\affiliation{$^2$Berkeley Center for Theoretical Physics, University of California, Berkeley, CA, 94720}
\affiliation{$^3$Theoretical Physics Group, Lawrence Berkeley National Laboratory, Berkeley, CA, 94720 }
\affiliation{$^4$Department of Physics, Stanford University, Stanford, CA 94305, USA}
\affiliation{$^5$Racah Institute of Physics, The Hebrew University, Jerusalem 91904, Israel}


\date{\today}

\begin{abstract}
We study the graviton phenomenology of TeV Little String Theory by exploiting its holographic gravity dual five-dimensional theory. This dual corresponds to a linear dilaton background with a large bulk that constrains the Standard Model fields on the boundary of space. The linear dilaton geometry produces a unique Kaluza-Klein graviton spectrum that exhibits a $\sim$ TeV mass gap followed by a near continuum of narrow resonances that are separated from each other by only $\sim 30$ GeV. Resonant production of these particles at the LHC is the signature of this framework that distinguishes it from large extra dimensions where the KK states are almost a continuum with no mass gap, and warped models where the states are separated by a TeV.
\end{abstract}

\maketitle


{\it{Introduction.}}
String theory is the leading candidate for a theory of quantum gravity. Its mathematical consistency dictates the existence of extra dimensions. However, the fundamental string scale, $M_s$, as well as the size of the extra dimensions are not necessarily tied to the four dimensional Planck mass, $M_{Pl}$, and string theory may become relevant at much lower energies~\cite{Antoniadis:1990ew, Witten:1996mz}. This opens the exciting possibility that the string scale is at TeV and string theory provides a solution to the hierarchy problem. The string theoretic relation:
\bea
M_{Pl}^2=\frac{1}{g_s^2} M_s^8 V_6
\eea
-- where $g_s$ is the string coupling, and $V_6$ the six-dimensional internal volume -- suggests two distinct scenaria. One is the well known AADD framework of large extra dimensions~\cite{Antoniadis:1998ig}, where the enormity of the Planck mass is accounted for by the large volume of the extra dimensions where gravity propagates and thus becomes weak, while the Standard Model interactions are localized effectively in our three spatial dimensions~\cite{ArkaniHamed:1998rs}.

The other possibility arises when both the string scale as well as the size of the extra dimensions are at a TeV~\cite{Antoniadis:2001sw}. In this case, the weakness of gravity is attributed to the smallness of $g_s$~\cite{Antoniadis:1999rm}. The hierarchy problem is now equivalent to understanding the smallness of $g_s$. A non-trivial limit of zero string coupling in string theory gives rise to a class of theories known as Little String Theories (LSTs) where gravity is completely decoupled~\cite{Berkooz:1997cq,Seiberg:1997zk} (see e.g.~\cite{Aharony:1999ks,Kutasov:2001uf} for review
 and~\cite{Aharony:2004xn} and references therein for a more recent work). On the other hand, gauge couplings, determined by geometric moduli, are independent of $g_s$  and can thus be order one.

LSTs can be generated by stacks of NS5-branes, similar to the way stacks of D3-branes may give rise to the Randall-Sundrum (RS) scenario ~\cite{Randall:1999ee, Verlinde:1999fy}. More precisely, in the gravity decoupling limit, D3-branes generate four dimensional (4d) ${\cal N}=4$ supersymmetric gauge {\it field} theories, while NS5-branes give rise to 6d LSTs which are notoriously hard to study; they are strongly coupled non-local theories that appear to have no Lagrangian description. Holography though allows to study n-dimensional theories without gravity in the strongly coupled regime by weakly coupled dual theories of gravity embedded in n+1 dimensions \cite{Maldacena:1997re}. In the well known RS case, the gravity dual is a 5d anti-de-Sitter background (AdS$_5$), while for LSTs one obtains a 7d theory with a linear dilaton background configuration in the infinite extra dimension~\cite{Aharony:1998ub}. It turns out that the level of difficulty is reversed in the dual theories: string theory on AdS background is very hard to study, while the linear dilaton has an exact world-sheet description. Thus, in contrast to RS scenario whose string embedding is challenging, LST models have in principle a well defined string realization.

In order to realistically associate the LST framework with the hierarchy problem and have a finite Planck scale, the infinite extra dimension of the dual theory needs to be rendered finite and the additional two transverse dimensions need to be compactified. This gives rise to a cigar type throat connected to an asymptotically flat spacetime. In this letter, we use a 5d analog of this geometry that nicely captures many of the LST properties. We discuss the graviton Kaluza-Klein (KK) spectrum and the possibility of having ordinary particles in the bulk. We finally discuss the collider phenomenology of this framework and how it can be distinguished from other possibilities.

{\it{The 5d model.}} As discussed above, the gravity dual of Little String Theory can be approximated by the following action in the bulk:
\bea
S_{bulk}=\int d^5x\sqrt{-g} e^{-\frac{\Phi}{M_5^{3/2}}} (M_5^3 R +(\nabla \Phi)^2 - \Lambda)
\eea
$\Phi$ is the dilaton field and $M_5$ is the five-dimensional cutoff, where bulk gravity becomes strong and is of the order of the string scale in the fundamental theory. The extra dimension is finite and compactified on an interval, described by a circle with a $Z_2$ symmetry.
We identify the boundary at $x_5\equiv y=0$ with the Standard Model (visible sector) brane, while at $y=r_c$ there is a hidden sector brane, and the corresponding actions are, respectively:
\bea
S_{vis(hid)}=\int d^4x \sqrt{-g} e^{-\frac{\Phi}{M_5^{3/2}}} \left(L_{SM(hid)}-V_{vis(hid)}\right)
\eea
In order to study the properties of this setup we can go to the Einstein frame, where the curvature term no longer has a dilaton field dependence. This is achieved by the conformal transformation:
\bea
\tilde g = e^{-\frac{2}{3}\frac{\Phi}{M_5^{3/2}}} g
\eea
and the above actions are rewritten as:
\bea
S_{bulk}=\int d^5x\sqrt{-\tilde g}  \left(M_5^3 \tilde R - \frac{1}{3}(\tilde\nabla \Phi)^2 - e^{\frac{2}{3}\frac{\Phi}{M_5^{3/2}}} \Lambda\right)\\
S_{vis(hid)}=\int d^4x \sqrt{-\tilde g} e^{\frac{1}{3} \frac{\Phi}{M_5^{3/2}}} \left({\tilde L}_{SM(hid)}-V_{vis(hid)}\right)
\eea
The LST solution arises when we impose a linear dilaton background $\frac{\Phi}{M_5^{3/2}}= \alpha |y|$. The gravity equations of motion in this background are solved by the following bulk metric:
\bea
ds^2=e^{-\frac{2}{3}\alpha |y|} \left(\eta_{\mu \nu}dx^\mu dx^\nu+ dy^2\right)
\eea
with the following conditions:
\bea
\Lambda=-M_5^3 \alpha^2~\text{and}~V_{vis}=-V_{hid}=4\alpha M_5^3
\eea
Similar to the case for RS, there are two tuning conditions. One is needed to cancel our 4d cosmological constant and the other results in the tuning of the radion potential. This second tuning becomes irrelevant after a stabilization mechanism is taken into account.

The Planck scale is determined by the size of the extra dimension, the slope of the dilaton field and the 5d cutoff:
\bea
M_{Pl}^2= 2 \int_0^{r_c} dy\, e^{- \alpha |y|} M_5^3= -2 \frac{M_5^3}{\alpha}\left(e^{-\alpha r_c}-1\right)
\eea
Given that the cutoff is of around the TeV scale it is first obvious that $\alpha <0$. In addition, this relation shows that gravity decouples when $r_c\rightarrow \infty$, as expected from the LST picture.
The sign of $\alpha$ is indeed compatible with the relation between our model and LST:
one may think about the NS5-branes as being located at $y=0$, where the string coupling ${\rm exp}(\Phi/M_5^{3/2})$ is large,
while the asymptotically flat regime away from the branes is at $y=r_c$, where $g_s^2=e^{\alpha r_c}\ll 1$.
Moreover, $\alpha$ and $M_5$ are related to the other
two parameters of the dual LST on the stack of NS5-branes, $M_s$ and the number of branes $N$, by:
\bea
\alpha=-\frac{M_s}{\sqrt N}~, \qquad M_5^3\simeq\frac{M_s^9V_6}{\sqrt N}~.
\eea

In order to develop a better intuition for this framework it is worth rewriting the metric in a form that is familiar for RS geometries using the coordinate transformation $dz=e^{-\frac{1}{3} \alpha y} dy$:
\bea
\label{physmetric}
ds^2_{LST}=\Big(1+\frac{|\alpha z|}{3}\Big)^2 \eta_{\mu \nu} dx^\mu dx^\nu + dz^2
\eea
We see that the gravity dual of LST is a geometry with power law warping and logarithmically varying dilaton. Thus, when we impose the constraint for the value of $M_{Planck}$ given the TeV scale cutoff, we find that the proper length of the extra dimension is quite large in fundamental units, $z_0~\approx (100~ {\rm eV})^{-1}(\sim 10~{\rm nm})$. In order to understand the physical implications of this dimension, it is essential to calculate the graviton KK spectrum.

{\it{Graviton Kaluza-Klein Modes.}} We find the spectrum of spin-2 excitations of the graviton $h^{(n)}_{\mu \nu}$ by taking the following parametrization of the metric:
\bea
e^{-\frac{2}{3} \alpha |y|}\left[(\eta_{\mu \nu}+h^{(n)}_{\mu \nu})dx^\mu dx^\nu+dy^2\right]
\eea
Working in the transverse-traceless gauge, we find that the equation of motion for these modes is:
\bea
\eta^{\rho \kappa} \partial_{\rho} \partial_{\kappa} h^{(n)}_{\mu \nu} + \partial_{y}^2 h^{(n)}_{\mu \nu} -\alpha \partial_y h^{(n)}_{\mu \nu}=0
\eea
The Neumann boundary conditions imposed by the symmetry result in a massless mode that is flat in the extra dimension, while the rest of the KK modes have wave-function localized close to the Standard Model brane:
\bea
h^{(n)}_{\mu \nu}=
N_{\mu\nu}^{(n)} e^{\frac{\alpha}{2} |y|}\! \left(\!- \frac{2 n \pi}{\alpha r_c}
\cos \frac{n \pi |y|}{r_c}+\sin\frac{n \pi |y|}{r_c}\right)e^{ipx}
\eea
where $N_{\mu\nu}^{(n)}$ accounts for a normalization factor and the tensor indices.
Their mass is given by:
\beq
m_n^2=\left( \frac{n\pi}{r_c}\right)^2 + \frac{\alpha^2}{4}\quad; \quad n=\pm 1,\pm 2,\dots
\eeq
We see that this metric gives rise to a quite special spectrum -- there is a mass gap of order the curvature scale followed by what essentially is a continuum of modes. This type of spectrum has been pointed out before in the LST literature (see e.g.~\cite{Aharony:1999ks}). This behavior of the geometry can be understood by rewriting the equation of motion for these modes as a Schroedinger equation with a potential by defining $h^{(n)}_{\mu \nu}=e^{\frac{\alpha}{2} |y|} \tilde h^{(n)}_{\mu \nu}$ :
\beq
\eta^{\rho \kappa} \partial_{\rho} \partial_{\kappa} \tilde h^{(n)}_{\mu \nu} + \partial_{y}^2 \tilde h^{(n)}_{\mu \nu}- \frac{\alpha^2}{4} \tilde h^{(n)}_{\mu \nu}=0
\eeq
This shows that the metric provides a bulk mass for the graviton modes, while only one remains massless because of symmetry. Above the mass gap, the spectrum is quantized as we would expect for a particle moving in a box of size $r_c \sim (30~\mbox{GeV})^{-1}$. This value comes in stark contrast with the proper length of the extra dimension, $z_o\sim 10 ~{\rm nm}$, as seeing from Eq.~(\ref{physmetric}). The puzzle is solved once we calculate the time it takes for a massless particle in this geometry to travel from one brane to the other, which is nothing else but $\sim (30~\mbox{GeV})^{-1}$ -- this sets the size of the ``box" in which the gravitons propagate.

In addition to the peculiarity of the mass spectrum, we expect the couplings of these excitations to be suppressed only by the TeV scale and not by the Planck mass, as they are localized close to the IR brane in our geometry. As a result, these modes can be produced and studied at colliders providing a smoking gun of our framework. We will discuss in more detail the couplings and the phenomenology of these modes below.
Finally, note that just as in the case of RS, there are no vector KK modes from gravity due to the $Z_2$ symmetry of the bulk.

{\it{KK graviton phenomenology.}} We previously showed that the KK excitations of the graviton are localized close to the SM brane and thus couple with a strength much larger than $M_{Pl}^{-1}$. Their exact couplings, $\frac{1}{\Lambda_\pi^{(n)}}$, are determined by the wavefunction normalization:
\bea
\int dy \left|\tilde h^{(n)}_{\mu \nu} \tilde h^{(n)\mu \nu}\right| = 1
\eea
This condition gives:
\bea
\label{KKcoupling}
\frac{1}{\Lambda_\pi^{(n)}}=\frac{1}{M_5} \bigg( \frac{|\alpha |}{M_5}\bigg)^{1/2}\frac{1}{|\alpha r_c|^{1/2}}\bigg(\frac{4 n^2 \pi^2}{4 n^2 \pi^2+ (\alpha r_c)^2}\bigg)^{1/2}
\eea
\begin{figure}[!t]
\begin{center}
\includegraphics[width=3.2in]{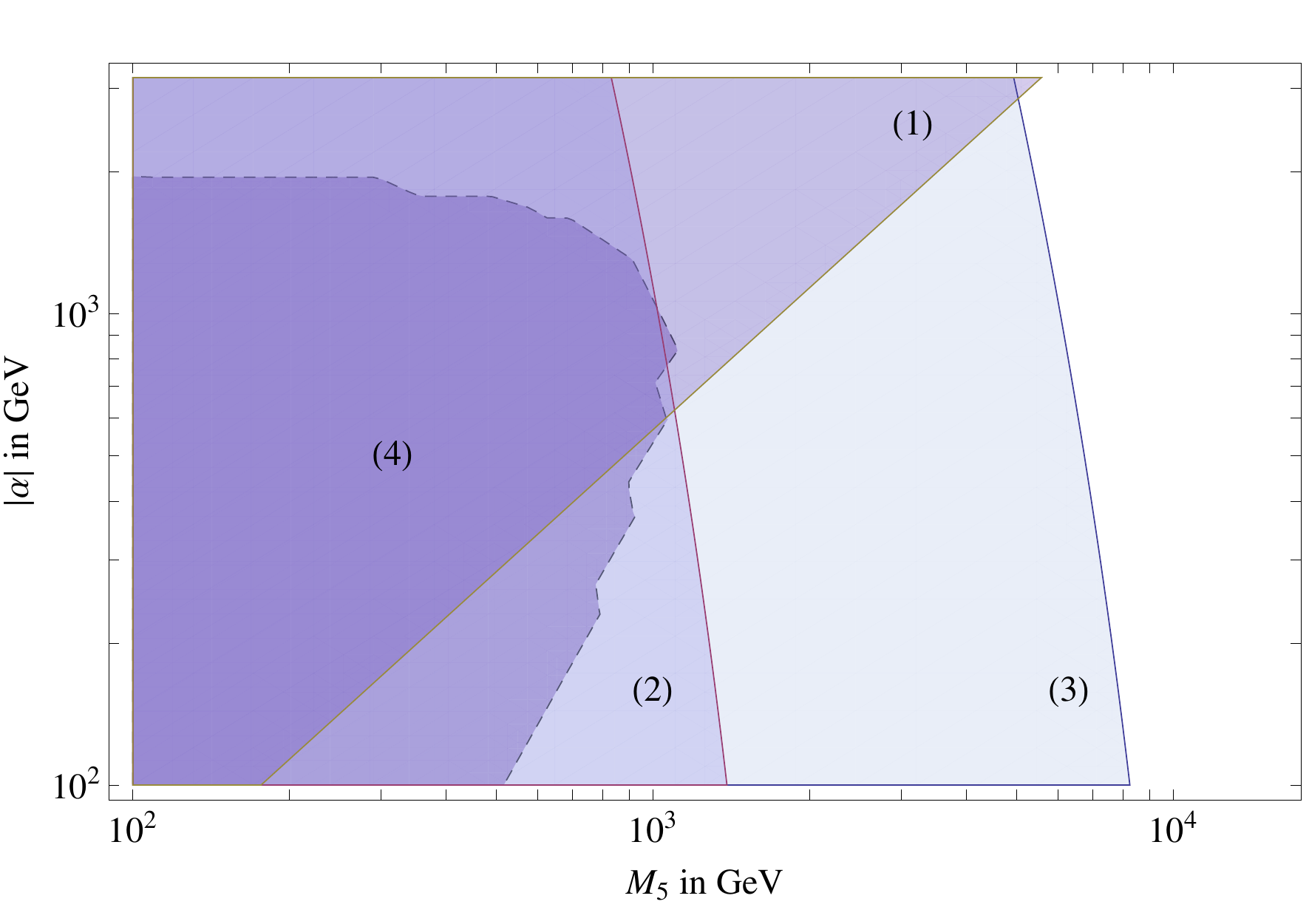}
\end{center}
\caption{Bounds on the LST parameter space from off-shell exchange of KK gravitons at Tevatron and the reach at the LHC. Starting from the darkest shade of blue (gray), the shaded regions represent the part of the parameter space (1) where the 5d curvature ${28\over 9}\alpha^2$
is larger than $M_5^2$, (2) that is excluded by the Tevatron with 5.4 $fb^{-1}$ of data, and (3) is accessible to a 14 TeV LHC with 10 $fb^{-1}$ of data. Region (4) contoured by the dashed line is the region excluded by direct searches in diphoton events at the Tevatron for 5.4  $fb^{-1}$ of data.}
\label{bounds}
\end{figure}
Eq.~(\ref{KKcoupling}) shows that each one of these modes couples with strength slightly smaller than $M_5^{-1}$. There is a factor of roughly 30 suppression coming from $\alpha r_c$, which needs to be large to get the correct value of $M_{Pl}$ as well as from the requirement of perturbativity $|\alpha|< M_5$. It is also worth pointing out that the total inclusive production cross-section of these states saturates unitarity at energies that are close to $M_5$, as expected.

Because of their relatively large coupling, these modes can be produced at colliders and appear as resonances in dilepton and dijet events. The search strategies are very similar to those for RS gravitons and the most stringent constraint currently comes from their off-shell production at the Tevatron \cite{Davoudiasl:1999jd}. Their contribution to contact interactions depends on the convergent sum:
\bea
\sum_n \frac{1}{{\Lambda_\pi^{(n)2}}} \frac{1}{s-m_n^2}
\eea

In Fig.~\ref{bounds} we present the current Tevatron bounds from contact interactions as we extrapolated from RS studies \cite{Davoudiasl:1999jd}, assuming $s < m_n^2$. We also show the expected reach at the LHC for these interactions. In the same figure we also present the current bounds from resonant KK production in the diphoton channel coming from the Tevatron with 5.4 $fb^{-1}$ of data. The direct bounds have been extrapolated from current RS searches \cite{Aaltonen:2010cf} and evaluated for different values of the KK number. In Fig.~\ref{bounds}, the resulting exclusion regions for different KK numbers have been over-imposed producing the dashed contour. As Eq.~(\ref{KKcoupling}) shows, it is a higher $n$-number KK mode that will be primarily produced  and set the most stringent bounds, since the KK coupling increases with $n$ and the KK states are closely-separated in mass. Fig.~\ref{bounds} should thus be taken as approximate and further study is needed to get the precise bounds.

It is obvious that there is plenty of parameter space to be explored by the LHC. The KK modes may be light enough to be directly produced at the LHC and, given the small mass splitting between the modes, several of them can be simultaneously accessible. In addition, the width of the KK states is much smaller than the mass splitting due to the coupling~(\ref{KKcoupling}). Thus, with a good energy resolution at the LHC, they can appear as distinct resonant peaks, making the mass relation predicted by this framework directly testable.

It is worth considering what happens to the allowed parameter space as $\alpha$ goes to zero. In the case of Fig.~\ref{bounds}, the assumption $s < m_n^2$ no longer holds, the KK gravitons are very light and the bounds can be approximated by those for AADD models. Astrophysical bounds also start becoming important when $m_n$ is less than roughly 100 MeV, and when $m_n$ is smaller than $10^{-3}$ eV, i.e. $\alpha < 10^{-3}$ eV, fifth force experiments already exclude this scenario.

{\it{Particles in the bulk.}} An important phenomenological feature of the setup is the behaviour of ordinary particles in the bulk. In order to see if the Standard Model can be removed from the boundary, we need to consider bulk gauge bosons. For these fields it is straight forward to see from the equation of motion (choosing $A_5=\eta^{\mu \nu} \partial_\nu A_\mu$=0):
\bea
\partial_M e^{-\frac{1}{3} \alpha |y|} \eta^{MN} \eta^{AB} F_{NB}=0,
\eea
that the zero mode has to be flat in the extra dimension, as expected by gauge invariance. The normalization of the mode in this metric is given by:
\bea
N^{-2}=-\frac{6}{\alpha}\left(e^{-\frac{1}{3}\alpha r_c}-1\right).
\eea
The corresponding effective 4d gauge coupling becomes so small that forbids putting Standard Model gauge fields in this bulk.

{\it{Discussion.}}  In our discussions we neglected to examine a very important aspect of the 5d LST framework: stability and the properties of the radion. It is easy to find the relevant modes by using the analysis in \cite{Kofman:2004tk}. As presented, our setup has two massless modes from the radion and dilaton fields as well as a bulk size that has nowhere been dynamically determined. But we do not expect this to be a problem; an analog of the Goldberger-Wise mechanism \cite{Goldberger:1999uk} for this framework should be enough to stabilize the size and give mass to any massless modes without a large back-reaction to the geometry. In this case, the dilaton itself is the GW field, after its value on the boundary is fixed \cite{Kofman:2004tk}. The dilaton potential can be stabilized with SUSY breaking, giving rise to what is known as a racetrack potential. The phenomenology of the radion deserves further study.

Moreover, our toy model does not capture all the features of ``LST at a TeV''~\cite{Antoniadis:2001sw}
such as perturbative fundamental string excitations, little string excitations,
KK modes of the two compact dimensions along the NS5-branes world-volume, as well as the KK excitations associated with the angular directions of the cigar geometry, that are all part of the full dual string theory description. The latter are actually the most relevant ones at low energies,
since they appear at a scale of order $|\alpha|$, together with the mass gap we discussed here.

In this work, we have seen that the LST phenomenology has certain characteristics of the more well studied frameworks of AADD and RS. It is a warped geometry, like RS, but with a large bulk that confines SM fields on the brane, like AADD. It is these characteristics that give rise to a unique KK graviton spectrum that can be discovered at the LHC.

{\it{Acknowledgements.}} We would like to thank Nima Arkani-Hamed, Gia Dvali, Sergei Dubovsky, Tony Gherghetta, Giovanni Villadoro, and Tomer Volansky for extremely useful discussions. Work supported by the European Commission under the ERC Advanced Grants 226371 and 228169, the contract PITN-GA-2009-237920, the NSF grant PHY-0756174, and the CNRS PICS no. 3747 and 4172.
The work of AG was supported in part by BSF, ISF (grant number 1665/10), and DIP grant H.52.
AG is grateful to the Theory Unit at CERN,
where this work was initiated, for its very warm hospitality.

\end{document}